\documentclass[
    ,final            
  ]
  {aipproc}

\layoutstyle{6x9}

\begin{document}

\title{Testing Observational Techniques with 3D MHD Jets in Clusters}

\classification{95.30.-k,95.30.Qd}
\keywords      {Magnetohydrodynamics, Active Galaxies, Galaxy Clusters}

\author{Peter J. Mendygral}{
  address={Department of Astronomy, University of Minnesota}
}

\author{Sean M. O'Neill}{
  address={Department of Astronomy, University of Maryland}
}

\author{Tom W. Jones}{
  address={Department of Astronomy, University of Minnesota}
}

\begin{abstract}
Observations of X-ray cavities formed by powerful jets from AGN in galaxy cluster cores are commonly used to estimate the mechanical
 luminosity of these sources.  We test the reliability of observationally measuring this power with synthetic X-ray 
observations of 3-D MHD simulations of jets in a galaxy cluster environment.  We address the role that factors such as jet intermittency 
and orientation of the jets on the sky have on the reliability of observational measurements of cavity enthalpy and age.  An estimate of the errors
in these quantities can be made by directly comparing ``observationally'' derived values with values from the simulations.  In our tests,
cavity enthalpy, age and mechanical luminosity derived from observations are within a factor of two of the simulation values. 
\end{abstract}

\maketitle


\section{Introduction}

X-ray images of giant cavities in galaxy clusters associated with powerful jets from central active galactic nuclei (AGN) suggest that
AGN may play an important role in the energetics of galaxy clusters (e.g. \cite{wise07,birzan04}).  
The minimum energy required to produce cavities, the enthalpy, is often in the range of $10^{55}$ to $10^{60}$ ergs
\citep{birzan04}.  The energy required to suppress the radiant cooling of the intracluster medium (ICM) below 2 keV is on the same order as 
energy in these cavities \citep{mcnamara07}.  One popular hypothesis that has emerged to solve this ``cooling flow problem'' 
is that AGNs release sufficient energy into the ICM to quench cooling.  A common method for estimating the energy released by an AGN requires measuring
the cavity enthalpy and age \citep{birzan04}.  We have produced synthetic X-ray observations from a pair of 3D magnetohydrodynamic (MHD) simulations 
of jets in realistic cluster environments.  The two simulations tested the observational implications of jet intermittency on cavities.

\section{Simulations}

We present two simulations of jets in galaxy clusters representing two models of jet intermittency \citep{oj}.  The simulations were computed 
using a 3D MHD total variation diminishing (TVD) code that employs flux constrained transport 
to maintain divergence-free magnetic fields (\cite{rj,ryu}).  The calculations were done on a Cartesian grid 600
zones in $x$ and 480 zones in $y$ and $z$ with $\Delta$x=$\Delta$y=$\Delta$z = 1 kpc.  Passive relativistic electrons were included following 
the Coarse-Grained finite Momentum Volume scheme \citep{treg01}.

The calculations were both initialized with an ambient environment intended to mimic many characteristics of observed galaxy clusters.
A static gravitational potential was included defined by an NFW dark matter distribution \citep{navarro} with a virial mass of 
4.2 $\times$ 10$^{14} M_{\odot}$ for a virial radius of 2 Mpc and core radius of 400 kpc.  The ICM was set in hydrostatic equilibrium 
with a core pressure of 2.4 $\times$ 10$^{-10}$ dyne cm$^{-2}$ and core density of 8.33 $\times$ 10$^{-26}$ g cm$^{-3}$.  A 
Kolmogorov spectrum of density fluctuations with an amplitude of $\sim$10\% was added on top of the initial density profile.  
The ICM magnetic field was $\sim$10 $\mu$G and tangled on scales of 
$\sim$10 kpc.  A pair of bi-directed jets originated from the center of the computational domain.  The jets were
launched in pressure equilibrium with the surrounding ICM at Mach 30 with respect to the ICM sound speed, $v_{jet}\sim$0.1 c. 
The jet diameter was 6 kpc .  Jet plasma was injected onto the grid with a density contrast of 
$\rho_{ICM}/\rho_{jet}$ = 100 compared to the core ICM density.  The jets contained a $\sim$10 $\mu$G toroidal magnetic field.
Two models of jet intermittency are tested; a relic model (RE) of jets turned on for 26 Myr and then permanently turned off, and an
intermittent model (I13) of jets toggled on and off every 13 Myr.  Jet power in both cases was intended to represent typical FRII radio jets.

Synthetic X-ray observations of both simulations were calculated over a band from 0.75 to 10 keV as a post processing step using a
ray casting technique.  Both thermal bremsstrahlung emission and inverse Compton scattering of CMB photons are included in the emission model.
The computational domains were set at a luminosity distance of 240 Mpc to produce images with a resolution comparable to \emph{Chandra}.  The grids 
were rotated to several inclinations to test for potential projection effects on the analysis.


\begin{figure}
  \includegraphics[height=.26\textheight]{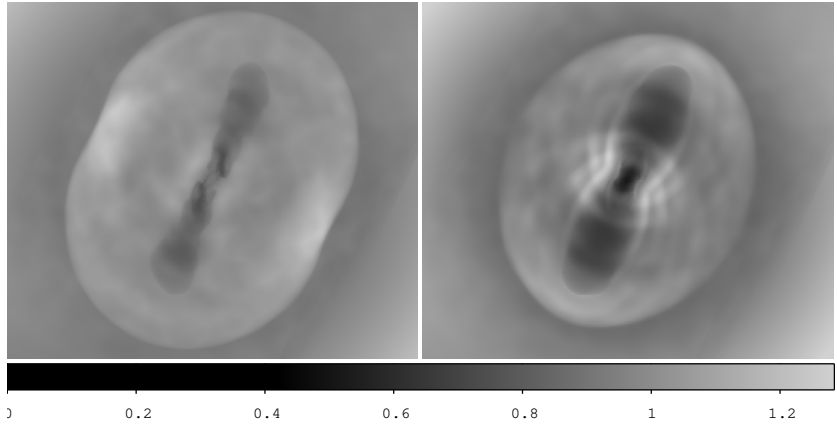}
  \caption{Synthetic X-ray observations of RE (\emph{left}) at 158 Myr and I13 (\emph{right}) at 171 Myr with the grids rotated to an inclination of
45$^o$ from 1.5 to 2.5 keV divided by a best fit double-$\beta$ profile to emphasize the cavities.  Note the presence of ripples in the I13 image.
These are produced as the jets are toggled on and off.  The I13 cavities have a wider aspect ratio as compared to RE.}
\end{figure}

\section{pV Work \& Mechanical Luminosity}

``Observational'' measurements of the minimum energy to produce the X-ray cavities visible in the synthetic X-ray observations were made following
a procedure similar to that of \citet{wise04} and \citet{rafferty06}.  X-ray cavities were fit by eye with ellipses to define their projected 
extent and volumes.  We then found best fit double-$\beta$ brightness profiles to X-ray images excluding the cavity regions.  Assuming an 
isothermal ICM, these profiles were inverted to produce an ICM density profile as given in \citet{xue00}.  With the assumption 
that the cavities were inflated in pressure equilibrium with the ICM, an estimate of the minimum energy required to produce the cavities, 
the enthalpy, is given by $H\,=\,\gamma{pV}/\left(\gamma\,-\,1\right)$.  Here $\gamma$ is 5/3 to match the simulation physics.

\begin{figure}
  \includegraphics[height=.32\textheight]{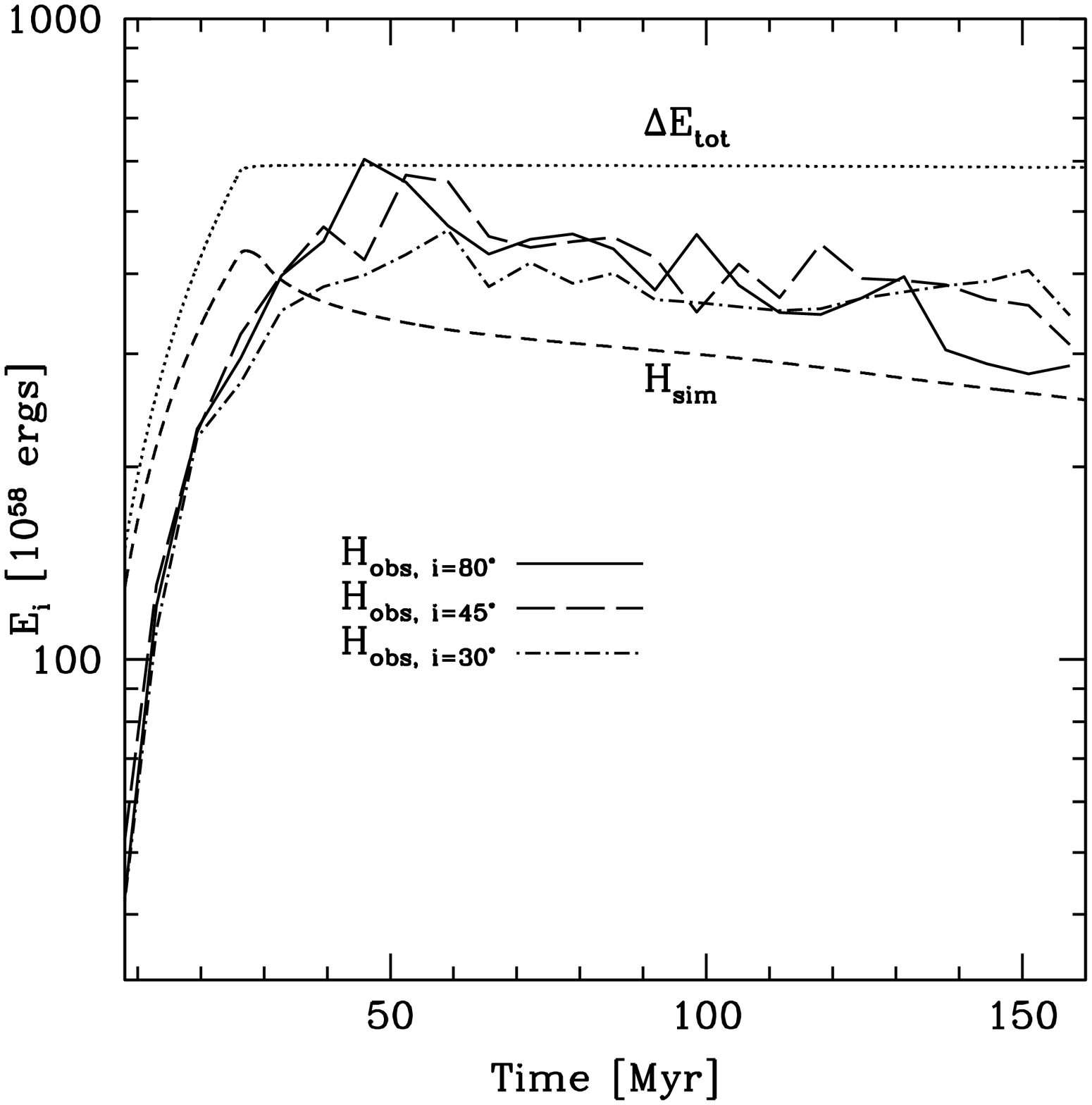}
  \includegraphics[height=.32\textheight]{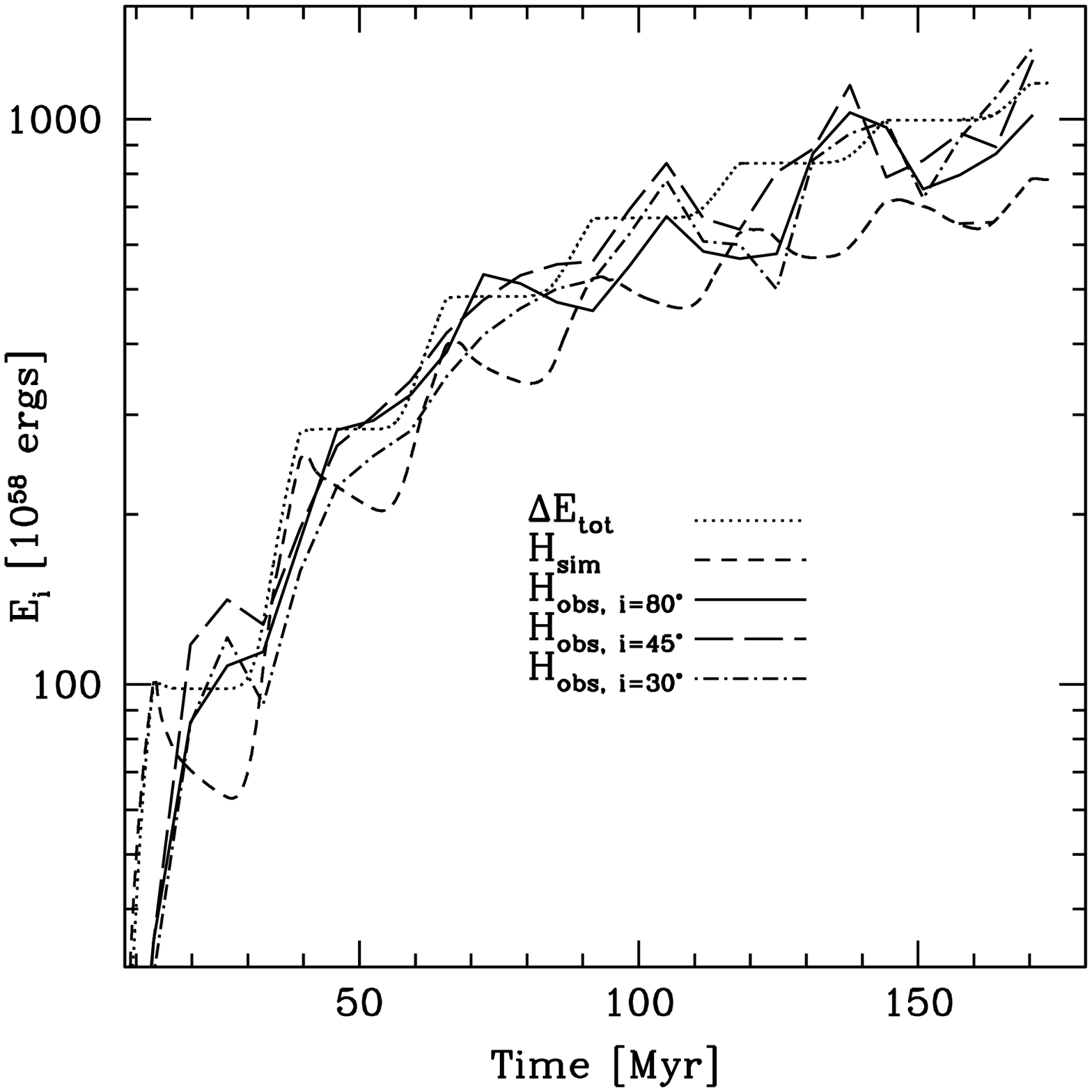}
  \caption{``Observed'' enthalpy, H$_{obs}$, for RE (\emph{left}) and I13 (\emph{right}) measured at various times through the simulation.  
Values are shown from observations made with the grid rotated to three inclinations.  The enthalpy measured directly from the simulation, H$_{sim}$, and
the total energy added to the grid from jets, $\Delta$E$_{tot}$, are shown for comparison.}
\label{fig:enth}
\end{figure}

Three commonly used cavity age estimates were calculated from the observations; buoyant rise time $t_{buoy}$, refill time $t_{r}$, sound crossing time 
$t_{c}$ \citep{mcnamara07}.  The mechanical luminosity of the jets was estimated from the enthalpy and an average of these three cavity ages as 
$L_{mech}$ = $H/\langle$t$\rangle$.

\begin{table}
\begin{tabular}{lllllllll}
\hline
  \tablehead{1}{l}{b}{Model\\}
  & \tablehead{1}{l}{b}{$i$\\}
  & \tablehead{1}{l}{b}{t$_{buoy}$\\(Myr)}
  & \tablehead{1}{l}{b}{t$_{r}$\\(Myr)}
  & \tablehead{1}{l}{b}{t$_{c}$\\(Myr)}
  & \tablehead{1}{l}{b}{$\langle$t$\rangle$\\(Myr)}
  & \tablehead{1}{l}{b}{H\\(10$^{60}$ ergs)}
  & \tablehead{1}{l}{b}{L$_{mech}$\\(10$^{44}$ ergs s$^{-1}$)}
  & \tablehead{1}{l}{b}{L$_{sim}$\\(10$^{44}$ ergs s$^{-1}$)}  \\
\hline
\hline
RE      & 80$^{o}$      & 103                   & 159           & 153           & 138           & 2.87                  & 6.59  & 12 \\
RE      & 45$^{o}$      & 89                    & 160           & 108           & 119           & 3.1                   & 8.25  & 12 \\
RE      & 30$^{o}$      & 71                    & 166           & 76            & 104           & 3.44                  & 10.5  & 12 \\
I13     & 80$^{o}$      & 89                    & 189           & 154           & 144           & 10.0                  & 22.0  & 21 \\
I13     & 45$^{o}$      & 67                    & 188           & 109           & 121           & 12.7                  & 33.2  & 21 \\
I13     & 30$^{o}$      & 54                    & 186           & 79            & 106           & 13.4                  & 40.0  & 21 \\
\hline
\end{tabular}
\caption{Several cavity age estimates, enthalpy and mechanical luminosity measured from ``observation'' for both RE and I13 at 158 Myr and 171 Myr respectively.
The jet luminosity in the simulation, $L_{sim}$, is given for comparison.}
\label{tab:a}
\end{table}

\section{Discussion}

The measurement of cavity enthalpy from the synthetic observations produced reasonable estimates compared to the actual enthalpy in the cavities in the simulation.
``Observed'' values tended to be larger than the simulation values by less than a factor of two for both RE and I13 as shown in Figure \ref{fig:enth}.  
Projection might have had a large effect on the 
accuracy of the measurements.  However, two factors counteract one another to produce only a small net variation in the measured values; cavities 
appear smaller and to be in a higher pressure environment closer to cluster core at small inclinations.
There was little or no dependence on the age of the cavities on the accuracy of the measured enthalpy.  This was largely due to the fact that cavities
closer to cluster core, such as young cavities, have a higher contrast ratio \citep{ensslin02} making it easier to define their extent.  The portion of older 
cavities in this regime contain the bulk of the total energy. Cavity ages from $t_{buoy}$ and $t_{c}$ varied significantly due to the variation in the projected 
distance from cluster center.  The ages estimated from $t_{r}$ have little dependence on orientation as the estimate relies only on a measurement of the 
gravitational acceleration and cavity radius.  $L_{mech}$ derived from the enthalpy and an average of the age values was within a factor of two of the real 
jet luminosity across models of intermittency and jet orientation as shown in Table \ref{tab:a}.  The energy required to offset cooling in clusters can be 
characterized as $\eta$pV \citep{birzan04}.  An approximate factor of two span in $\eta$ in our tests is due largely to uncertainties in the measurement of pV.


\begin{theacknowledgments}
This work has been supported by NSF grants AST06-07674 and AST09-08668 and by the University of Minnesota Supercomputing Institute.
\end{theacknowledgments}



\bibliographystyle{aipproc}   




\end{document}